# The Role of Elastic Stresses on Leaf Venation Morphogenesis

Maria F. Laguna[1]*, Steffen Bohn[2,3], Eduardo A. Jagla[1]

1 Centro Atómico Bariloche, Comisión Nacional de Energía Atómica, Bariloche, Argentina, 2 Matière et Systemes Complexes, UMR 7057 CNRS and Universitè Paris 7 - Denis Diderot, Paris, France, 3 Center for Studies in Physics and Biology, Rockefeller University, New York, New York, United States of America

**Abstract**

We explore the possible role of elastic mismatch between epidermis and mesophyll as a driving force for the development of leaf venation. The current prevalent 'canalization' hypothesis for the formation of veins claims that the transport of the hormone auxin out of the leaves triggers cell differentiation to form veins. Although there is evidence that auxin plays a fundamental role in vein formation, the simple canalization mechanism may not be enough to explain some features observed in the vascular system of leaves, in particular, the abundance of vein loops. We present a model based on the existence of mechanical instabilities that leads very naturally to hierarchical patterns with a large number of closed loops. When applied to the structure of high-order veins, the numerical results show the same qualitative features as actual venation patterns and, furthermore, have the same statistical properties. We argue that the agreement between actual and simulated patterns provides strong evidence for the role of mechanical effects on venation development.

**Funding:** MFL and EAJ are financially supported by Consejo Nacional de Investigaciones Científicas y Técnicas (CONICET), Argentina. Part of this work was performed under grant PICT-2005/32859 (ANPCyT, Argentina). SB is partly funded by Human Frontiers Science Program number 62/2005.

**Competing Interests:** The authors have declared that no competing interests exist.

* E-mail: lagunaf@cab.cnea.gov.ar

## Introduction

For many years leaf venation motifs have marveled people, whether scientists or not. Venation patterns are different from one leaf to another, even in the same plant, but share some common features that are preserved throughout all angiosperm leaves [1]. A remarkable characteristic of these patterns is the vein hierarchy, characterized by their radii, that originates in the successive formation of veins during leaf growth [2,3]. A second, very robust, feature of the venation pattern is the abundance of closed loops: the leaf surface is divided into small polygonal sectors by the venation array; only the fine veins of the highest orders do not connect at both ends and are often open ended (see Figure 1).

It has been argued that the vein architecture might ensure optimal water distribution [4,5]. However, the straightforward optimization of steady state irrigation within the leaf must lead to tree-like open topologies [4,5] with strictly no loops [6]. The high redundancy of paths from the leaf base to any point in the leaf surface might nevertheless be very advantageous with regard to local damages (Magnasco MO, personal communication). Also, it has been suggested that venation may play a mechanical stabilization role for the leaf, but the optimization of the mechanical stabilization leads to very unnatural venation geometries [5].

From a developmental perspective, the leaf venation is puzzling, too. Since the pioneering works of Sachs [7–9], it is known that the growth hormone auxin has an enormous effect on the venation pattern [10–12]. It is believed that auxin is synthesized in the growing leaf (either homogeneously or at localized sites) and that there is a net auxin flow towards the leaf base from where it is transported towards the plant roots. Furthermore, it has been found that mutations that affect the auxin transport lead to strongly modified venation patterns [13,14].

These findings have led to models of venation formation based on a positive canalization feedback [7–9,12]: on the one hand, the auxin flow is canalized into veins and vein precursors (procambium). On the other hand, high auxin concentrations (or, in a different variant, high transport values) trigger the differentiation into procambium. In its simple form, this model cannot lead to any loop but gives rise to tree-like structures [15–18], and this is a serious drawback of the model. Several studies have tried to correct this unrealistic part of the model with varying success [19–23]. For instance, Rolland-Lagan and Prusinkiewicz [20] have proposed the possibility that localized auxin sources on the leaf move around when veins develop. They show that closed loops can be formed in this way. This model seems to require a rather complex and coordinate displacement of auxin sources as veins are formed. On the other hand, Dimitrov and Zucker [21] have considered a homogeneous production of auxin on the surface of the leaf, and suggested that closed loops are formed when new vein segments propagate from existing ones, and meet at the point of highest auxin concentration. From a basic perspective, it seems that this model requires a very precise coordinated progression of the new vein segments, as otherwise the first segment reaching the highest auxin concentration point would inhibit further growth and open structures would be obtained. Along the same lines, Runions and collaborators have devised geometric algorithms that give rise to aesthetically very appealing venation patterns [22]. Closed loops are obtained in this case (as in [21]) by the tips of three vein segments meeting in points of high auxin concentration.

### Author Summary

Leaf venation patterns of most angiosperm plants are hierarchical structures that develop during leaf growth. A remarkable characteristic of these structures is the abundance of closed loops: the venation array divides the leaf surface into disconnected polygonal sectors. The initial vein generations are repetitive within the same species, while high-order vein generations are much more diverse but still show preserved statistical properties. The accepted view of vein formation is the auxin canalization hypothesis: a high flow of the hormone auxin triggers cell differentiation to form veins. Although the role of auxin in vein formation is well established, some issues are difficult to explain within this model, in particular, the abundance of loops of high-order veins. In this work, we explore the previously proposed idea that elastic stresses may play an important role in the development of venation patterns. This appealing hypothesis naturally explains the existence of hierarchical structures with abundant closed loops. To test whether it can sustain a quantitative comparison with actual venation patterns, we have developed and implemented a numerical model and statistically compare actual and simulated patterns. The overall similarity we found indicates that elastic stresses should be included in a complete description of leaf venation development.

Nevertheless, it is an open question whether the auxin sources postulated in [21,22] for the formation of high-order veins actually exist, since Scarpella *et al.* [12] failed to observe them in their experiments.

An alternative model has been recently introduced by Feugier and Iwasaa [17,23]. In this model, loops are formed when a vein tip curves towards and meets an older vein at some intermediate point. It is suggested that this behavior is induced by the existence of 'flux bifurcators' in some of the cells with high auxin concentration. Note that this mechanism is incompatible with the one proposed in [21,22], as here the loops close at intermediate points of older veins. Whether the hypothesis of Feugier and Iwasaa can generate realistic venation patterns is an open question.

In general, we find that the modifications to the canalization hypothesis necessary to explain the existence of closed loops are not generic and rather unnatural, and the mechanism on which they are based require a lot of fine tuning.

Couder et al. [24] have pointed out that the difficulties encountered in creating realistic, loop forming models on the basis of auxin transport are intrinsically related to the scalar nature of the concentration fields. In contrast, the growth in a tensorial field gives rise to hierarchical networks in a very robust manner. They suggested that this tensorial field could be the mechanical stress field in the growing leaf. (In a certain sense, the PIN protein polarization field in [23] can also be considered as a kind of tensorial field.) In their work they put forward the hypothesis that mesophyll cells that are submitted to compressive stress exceeding a threshold value start a differentiation process that eventually transform them into procambium. This process would be similar to the one observed in experiments on botanical tissues in which oriented cell divisions are forced by externally applied compressive stresses [25,26].

Evidence supporting this hypothesis is two-fold. On the one hand, micrographs taken in the early steps of leaf venation development show that in the first stages of differentiation, cells forming the procambium can be distinguished from the remaining cells by a mechanical distortion, consisting in a shrinkage of the cells perpendicular to the vein direction (see, for example, the images of Figure 2 of [2]). This suggests that stresses play a role in this distortion. On the other hand, it has been shown that typical large-scale morphologies of leaf venation patterns can be reproduced as crack patterns in an appropriately prepared layer of a slurry that dries in contact with a substrate [24]. This visual similarity between crack and venation patterns led us to investigate in more detail the fundamental ingredients in crack pattern appearance.

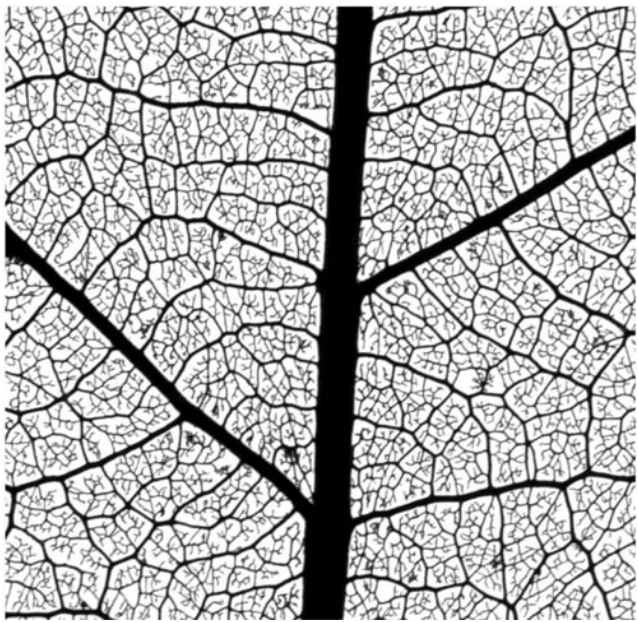

**Figure 1. Venation pattern of a *Gloeospermum sphaerocarpum* leaf.** This leaf was subjected to a chemical treatment to remove all the soft tissues, leaving only the veins. The network-like structure as well as many open ends of the thinnest segments can be observed.
doi:10.1371/journal.pcbi.1000055.g001

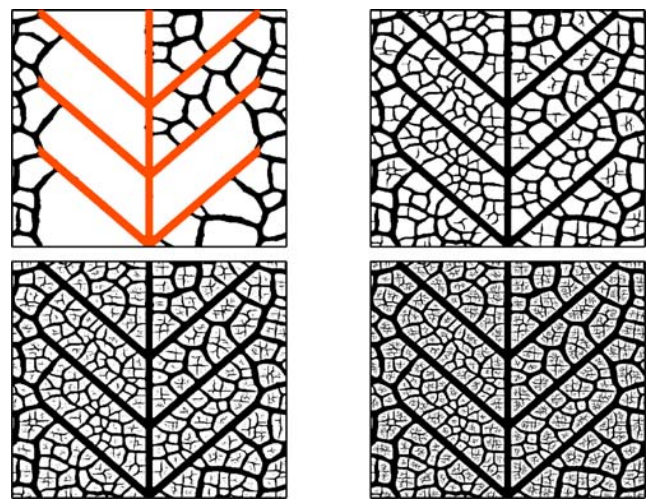

**Figure 2. Snapshots of the development process.** The values of the growing parameter, from top left to bottom right, are $\eta = 1.2$, 2.4, 3.6, and 4.8. The seed we use as the initial condition is shown in the first panel with a different color. The numerical lattice has $1024 \times 1024$ nodes.
doi:10.1371/journal.pcbi.1000055.g002

Crack patterns on the surface of mud or other materials require the existence of two quasi-two dimensional layers of material, the substrate and the covering, the latter contracting with respect to the former upon desiccation. (A pioneering work by Skjeltorp and Meakin [27] analyzes experimental and computer models of crack growth in a two-dimensional system consisting on two layers growing at different rates.) A rather similar situation may indeed occur in a growing leaf. In fact, a growing leaf consists of two epidermal layers separated by a softer tissue called mesophyll. This mechanical unit has to keep its integrity through the growing process. In the first stages of cellular growing and division, the three layers keep their status of uni-cellular layers. However, the growing rate of the epidermis and the mesophyll are not equal but the mesophyll tends to grow more rapidly than the epidermis [28]. This generates compressive stresses in the mesophyll that can force cells to grow and divide along particular directions, favored by the local stress field. In fact it is in this stage where evidence of collapsed cells of the mesophyll has been obtained [2]. We interpret the existence of elongated cells as evidence of a larger mechanical stress along the directions perpendicular to the largest axis of the deformed cells. Note that the similarity between crack patterns and our mechanical model for leaf venation has an important difference: crack patterns are obtained under contraction of the active layer relative to the substrate, whereas venation patterns should appear when there is an expanding active layer (mesophyll) relative to a rigid frame (the epidermis).

The suggestion of Couder et al. on the importance of elastic factors in vein formation [24] has not been further studied from a modeling point of view. In this paper, we present a numerical model based on this hypothesis. We will show that this approach, assuming the existence of a mechanical collapse instability of the mesophyll cells, generally leads to patterns that are not only qualitatively similar to actual venation patterns, but also show comparable statistical properties.

## Results

### Numerically Generated Patterns

In actual leaves, there is an obvious dependency between the morphology of veins and its rank in the venation structure. In other words, initial vein generations are strongly dependent on the form of the leaf and most probably, on genetic factors. It is this large-scale pattern that is repetitive within the same species and allows a broad leaf classification according to their venation patterns. It is also in these initial vein generations where the role of auxin is relatively well established. High order vein generations are much more isotropic, and much more universal in its statistical properties. It is to this stage that we intend to apply our model in its present form to compare statistical properties.

A comprehensive mathematical description of our model is given in the last section, but here we summarize the main hypotheses to ease the reading of this part. We assume that during growth, the inner cell layer (the mesophyll) is elastically attached to the epidermis. The epidermis is assumed to grow at a lower rate than mesophyll, and is otherwise supposed to be inert, i.e., it undergoes no deformations during growth. Due to the different growth rates of mesophyll and epidermis, compressive stresses develop in the mesophyll. Our main assumption is that the elastic properties of the mesophyll are such that this compressive stress can give rise to a shape change of the mesophyll cells. Such cells will acquire an elongated shape perpendicular to the main applied stress. These assumptions are basically equivalent to the description of collapsing surface layers presented in [29]. As in this work, the elastic properties of the mesophyll are included in the definition of a local free energy that has two minima: an isotropic 'intact' minimum, and a 'collapsed' one that corresponds to the deformed cell. (From a biological point of view, what we are describing as the 'collapse' of a cell from a rather spherical shape to an elongated shape could occur as preferential growth along the easiest direction, i.e., perpendicularly to the compressive stress field.) We use an algorithm in which the elasticity of the cells is assumed to be linear, the non-linear behavior is introduced by a scalar field $\Phi(x,y)$. The value of $\Phi(x,y)$ carries the full information of the complete tensorial stress field and the state of the system at the $(x,y)$ position. As will be clear in the last section, the field $\Phi$ has two preferred values, defining two elastic states with different density and shear modulus. They represent the intact and collapsed states of the cells in our model. Sectors of the system that are in the intact or collapsed states are recognized by their different values of $\Phi$ (see typical profiles of $\Phi$ in the last section). We will typically refer to collapsed sectors as 'veins', although it must be kept in mind that the definitive differentiation of a vein will require a further process that we are not modeling here. At each step of the simulation the system evolves towards the configuration that minimizes the total free energy. At the same time, a parameter $\eta$ (see the precise definition in the last section of this paper) is used to control the global growing of the leaf: increasing the parameter $\eta$ simulates the increasing of the overall leaf size. For technical simplicity we maintain the size of our simulation mesh (typically $1024 \times 1024$ nodes with periodic boundary conditions), and the increase in $\eta$ means that we are effectively 'zooming out' with the leaf growth. This means that new veins will be seen as thinner ones, while older veins keep their thickness during the simulation. In order to have a reasonable description of the hierarchical process of sequential vein formation, a sort of 'irreversibility' condition is implemented. It guarantees that once a new vein is created, it is forced to remain in the collapsed state during the leaf growth. In actual leaves, a similar mechanism explains why older veins are thicker: once a cell becomes a vein cell, the process of cellular division generates new cells that will also be vein cells. The implementation of the irreversibility condition in the model is explained in detail in the last section.

To avoid an extremely uniform initial condition, we typically seed the simulation with a few large-scale veins that provide the initial veins of our numerical leaf. This first division is not significant in the statistical analysis we perform on the final patterns. We show results in which we prepare the system with tree-like thick initial veins, or divide the sample into two pieces.

When new veins are formed (upon increasing of $\eta$), they typically propagate rapidly through the system, reaching in most (but not all) cases an older vein, where they stop. This propagation, once triggered, occurs essentially at constant $\eta$, i.e., it is not driven by the growing itself.

A few snapshots during the numerical evolution are shown in Figures 2 and 3, where we plot the points of the numerical mesh for which $\Phi$ have positive values (associated to the collapsed state). The hierarchical nature of the process can be clearly observed in these figures, as new veins are progressively thinner than old ones. We stress that the observed hierarchical patterns are a direct consequence of the irreversibility condition. In this way the history of the growth process remains encoded in the statistics of vein widths. Moreover, notice that hierarchical patterns can also be obtained in a very simple and well-controlled model such as that described in Text S1.

Before going to the quantitative characterization of the patterns obtained, two important features are worth noting. One is that in many cases several thin free-ended veins are observed. This also occurs in actual leaves and we propose an explanation in the next section. Another feature is that some minor veins are completely disconnected from other veins. They typically appear at the center of intact regions (where the stress is maximum), and seem unrealistic, since vein patterns in leaves are almost always connected. Although they might be due to an artifact in our simulations (in fact, the thickness of these disconnected veins is already comparable to our numerical discreteness), recall that our patterns are actually showing the places where the tension is high enough to generate collapsed cells that will eventually, but not necessarily, differentiate into veins. If the later differentiation process requires the canalization of a flux through the network of collapsed cells, differentiation of the disconnected segments into disconnected veins will not occur.

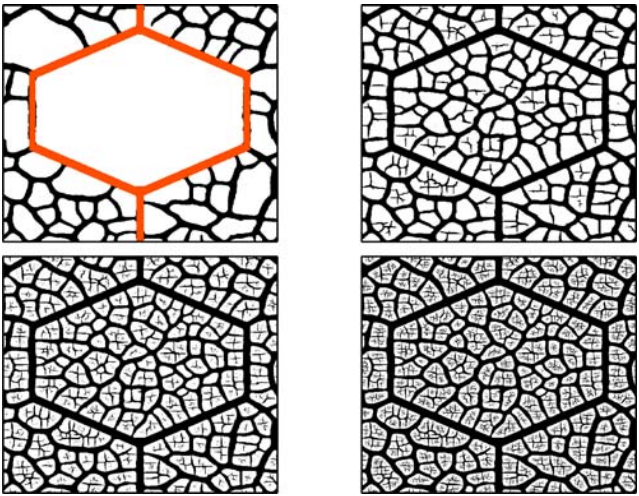

**Figure 3. Snapshots of the development process with a different starting configuration.** The values of η and the system size are the same as in the previous figure. In both figures the hierarchical process can be clearly observed. Note also the open ends of some of the thinnest segments.
doi:10.1371/journal.pcbi.1000055.g003

## Quantitative Comparison with Actual Leaf Patterns

In order to test whether our simulation results are comparable with actual leaf patterns, we computed the vein width, length and angles from our simulation results, and compare them with data from actual leaves. The same numerical image processing technique was used for the two data sets; see a detailed explanation in [30]. The image processing converts the venation patterns in sets of segments, nodes and free endings, each segment having a given length and width.

In Figure 4 we show the average length of the vein segments as a function of its width, $w$. Data from actual leaves of Figure 4A show that at first glance the typical length of segments is independent of the segment width, except for very thin segments, since there is a minimum thickness below which there are essentially no segments. This result is obtained also in the toy model presented in Text S1. An interesting deviation of this trend is found however when averaging many different data sets, where we see that thicker segments tend to be slightly longer than thinner ones (see the inset of Figure 4A). Going back to snapshots of actual leaves (Figure 1), it is clear that this result originates in the fact that thin segments have some difficulty in reaching thick segments and many open ends of thin segments are typically found near thick ones. Notably, this feature is reproduced in our numerical model (see Figures 2 and 3), and the increase of length as a function of segment width is in fact observed in the statistical plot of Figure 4B. The reason for the difficulty of thin segments to reach thicker ones in our model (and probably also in actual leaves) is the following. A given vein segment relaxes mechanical stresses in some neighborhood of it. The size of this relaxed zone increases with the vein width. When a thin vein is approaching a thick one, it enters a region where elastic stresses have diminished, and in many cases this relaxation is sufficient to stop the advance of the thin vein before it actually hits the thicker one. In case of approaching veins of approximately the same thickness this tendency is lower, and it does not seem to be strong enough to stop the vein advance before contact.

Moving to the description of the results of Figure 5 for the number of vein segments with a given width, $\mathcal{N}(w)$, first of all we note the overall similarity of real and numerical curves. Also, a shoulder in $\mathcal{N}(w)$ is observed both in the numerical as in the real data for the region of thick segments. In our numerical leaves we relate this behavior with the way in which we seed the simulation. In our runs, the first generation of veins appears quite rapidly and generates a number of thick segments. We observed that such distribution of thick veins is quite constant during the evolution of the system, whereas the region of the curve fitted by a power decay appears in later stages of the growing. The evolution of $\mathcal{N}(w)$ can be observed in Figure 6, where we plot the histograms of widths for the four snapshots of Figure 2.

For intermediate values of thickness, the results of our model are compatible with a power decay of $\mathcal{N}(w)$, with an exponent close to 2 (see Figure 5B). This result is also obtained with the minimal model described in Text S1, showing that our model generates a hierarchical pattern along the lines we have already discussed. From the data of actual leaves of Figure 5A we see that $\mathcal{N}(w)$ can be fitted by a power law decay, and this is a nice indication that a hierarchical mechanism is at work in actual leaves. However, in this case the decay exponent of $\mathcal{N}(w)$ is larger than 2, rather close to 3. Although it is probably too ambitious to try to give an explanation of this discrepancy, we want to present the following argument. One of the implicit assumptions in our scaling method is that all distances measured over the leaf surface grow at the same rate during leaf growth. This is reasonable as long as the cellular layers involved are one-cell thick. However, once some cells have been committed to become a vein, they must give rise to a cylindrical object. The hypothesis of two-dimensionality does not work for veins. If, on biological grounds, we assume that the rate of cellular division is constant, and take it independent of the kind of cell, we arrive to the conclusion that vein width increases as square root of time, instead of linearly. If this fact is taken into account in a counting as we did in the model described in Text S1, the result is that $\mathcal{N}(w)$ gets an additional factor $w^{-1}$, justifying a more rapid decay for $\mathcal{N}(w)$ in actual leaves than in our model, which assumes all distances measured in the leaf surface grow at the same rate.

Finally, we analyze the behavior of the angles between vein segments at the points where three vein segments meet. As pointed out in [30], the values of the three angles of a node are directly related to the local hierarchy of the meeting vein sizes. The authors found that the relation between angles and radii (or widths) is a general property of all the leaves they studied. We analyze our patterns to see whether it is possible to find in the numerical leaves the kind of organizational law obtained in actual venation patterns. For each node, we measure the three angles obtained and relate them with the radii of the vein segments. Thus, $\alpha_{LS}$ is the angle between the thickest and the thinnest segments, $\alpha_{LI}$ is the angle between thick and intermediate segments, and $\alpha_{IS}$ is the angle between intermediate and thin segments. We calculated the averages of the three angles and plot them as a function of the ratio between the radius of the thinnest ($R_S$) and thickest ($R_L$) segments. The configuration of radii is well defined with the parameter $R_S/R_L$ because the segment of intermediate radius has usually a value close to $R_L$. In Figure 7 we compare the numerical and the real data by adding our numerical results to the ones of Figure 14 of [30]. A very good agreement is obtained. The behavior observed can be understood by analyzing the two limiting cases. For $R_S/R_L$ close to one, all radii are almost equal and the three angles are near to 120 degrees. This describes a situation in which a vein has bifurcated into two. Since the three segments are then created almost simultaneously, the three radii are similar. On the other hand, $R_S/R_L$ near to zero correspond to the case in which a thin vein reaches a thick one. In this case, the angle $\alpha_{LI}$ between thick and intermediate segments tends to be 180 degrees, meaning that the thick vein is almost unperturbed by the thin one. A continuous and rather linear variation is observed between these two extreme situations. Although the overall coincidence of measured angles in our simulations and in actual leaves is encouraging, a full understanding of the origin of a general relation between angles and radii is not achieved yet.

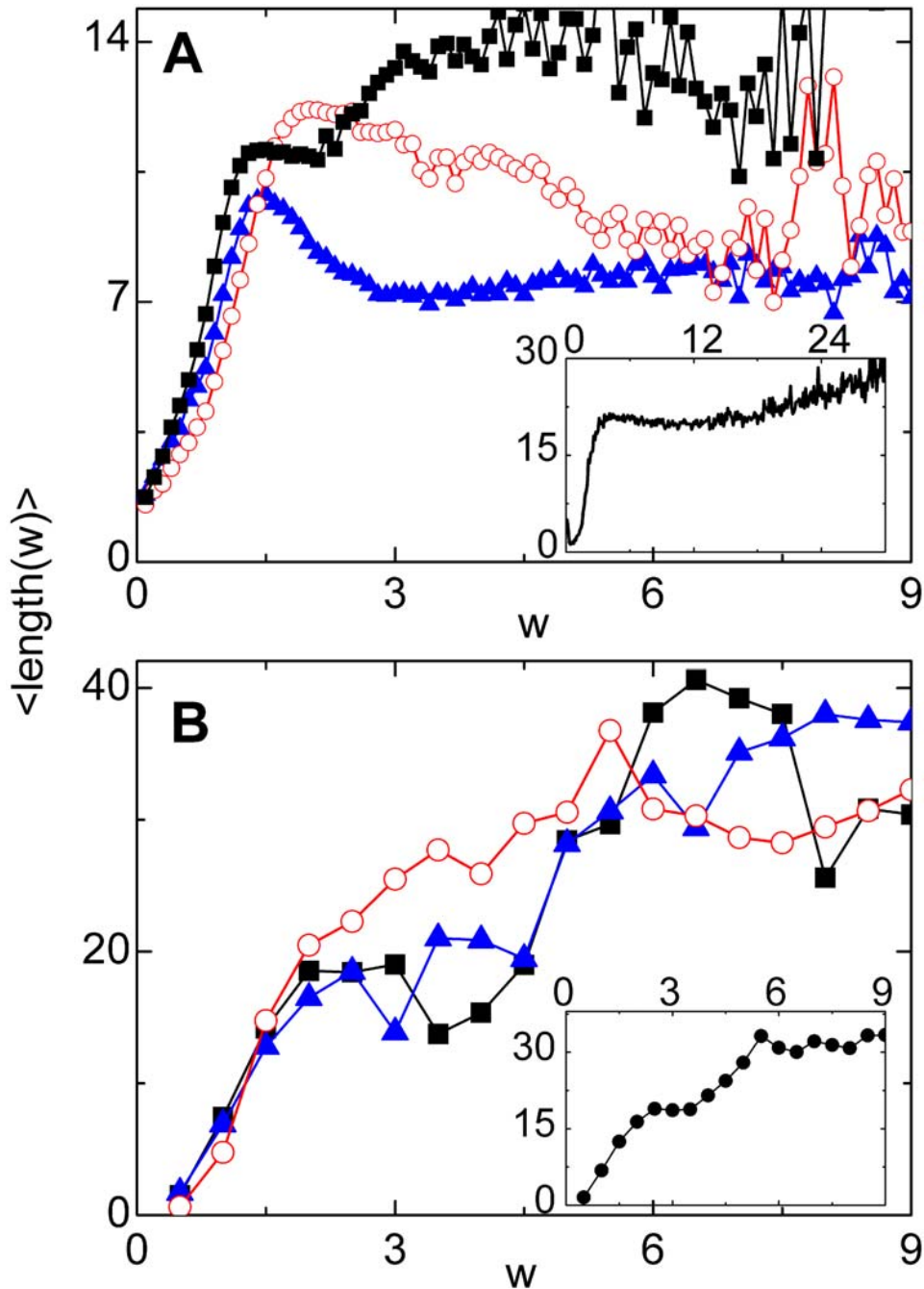

**Figure 4. Histograms of the average length of the vein segments of width *w*.** (A) Actual leaves. Each curve is the histogram of a given dycotiledon leaf: *Gloeospermum sphaerocarpum* (square symbols), *Amphirrhox longifolia* (full triangles), and *Rinorca amapensis* (open circles). Inset: the same quantity, but averaged over more than 1,200,000 segments of eight different leaves. Note that thicker veins tend to be slightly larger than thinner ones. (B) Numerical leaves. Histograms for three different realizations (size: 1024×1024). Inset: Histogram of 30,000 segments obtained in twelve realizations for $\eta = 3.6$ and three different sizes (512×512, 768×768, and 1024×1024).
doi:10.1371/journal.pcbi.1000055.g004

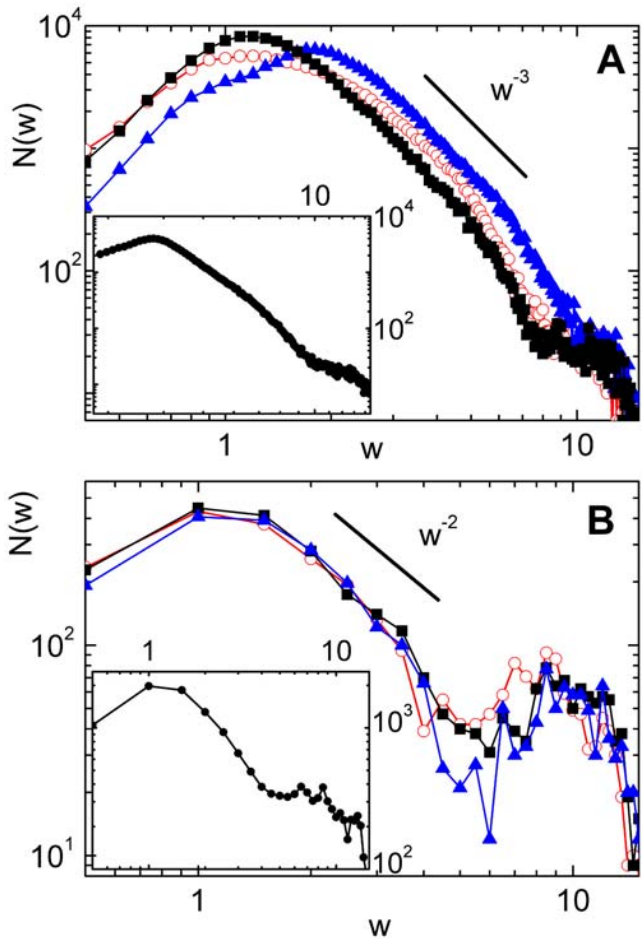

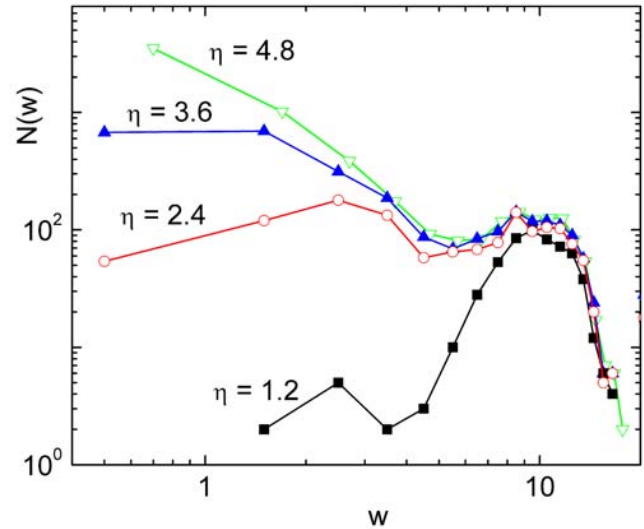

**Figure 6. Evolution of the histograms of widths.** Each curve corresponds to one of the four stages of growth shown in Figure 2 of this paper. Note that the distribution of thick veins is quite constant during the evolution.
doi:10.1371/journal.pcbi.1000055.g006

**Figure 5. Histograms of the number of vein segments of width $w$.** (A) Actual leaves. Histograms for the same three leaves showed in the previous figure. For all the leaves analyzed, a power decay with an exponent close to 3 is observed. Inset: Average over four leaves. A shoulder for thick veins can be observed in both figures. (B) Numerical leaves. Histograms for three different realizations. In the region of intermediate values of thickness, a power decay with an exponent close to 2 is obtained. Inset: Average for the same realizations as in the previous figure, showing a shoulder for the region of thick veins.
doi:10.1371/journal.pcbi.1000055.g005

In our model, the free energy of a vein can be conceived as a interface energy between the two sectors into which the vein divides the leaf. In the case that all veins are of the same width, the minimization of this interface energy would give rise to a foam-like pattern with 120 degrees angles. However, irreversibility gives rise to the formation of veins of different thickness and free energy minimization produces angles whose values are correlated with the veins' age.

The 'force model' proposed in [30] shows that if a force is assigned to each vein segment, pointing along the segment direction and with an intensity proportional to the vein radius, the angles between segments correspond to the situation in which the three forces emerging from each node are in equilibrium. The applicability of the force model to our numerical results could be justified by the following argument. Assuming that three segments of given radii have to meet, our modeling prescribes that the structure they form must have the minimum accessible free energy. If we assume that a rough measure of the free energy is given by the area covered by the veins, a line tension can be associated with each vein, which is proportional to its radius, and from here the prediction of the force model follows immediately. In any case, this is a point that deserves further study.

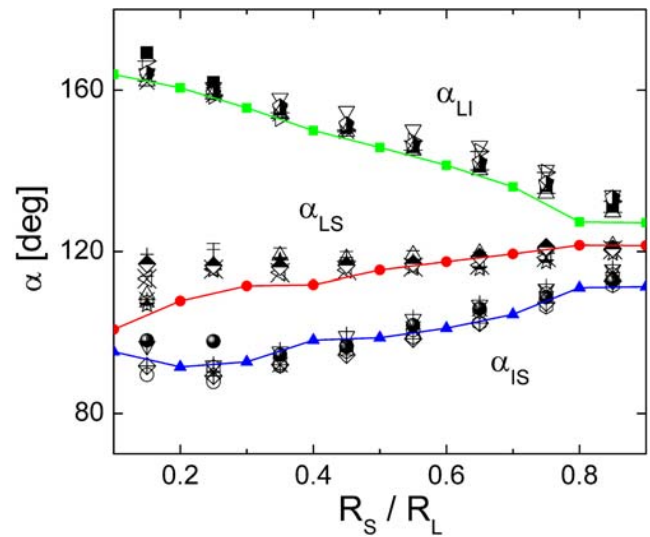

**Figure 7. Comparison of angles.** Angles between veins as a function of the ratio between the radius of the thinnest ($R_S$) and thickest ($R_L$) segments. The angle between thin and intermediate radius is labeled $\alpha_{IS}$. The angle between thin and thick segments is $\alpha_{LS}$, whereas the angle between thick and intermediate segments is $\alpha_{LI}$. Isolated symbols are data obtained from actual leaves, and were taken from Figure 14 of [30]. Colored lines with small symbols are our numerical results.

## Discussion

In this paper we have set up a model to study leaf venation, which is based on the idea that venation patterns are strongly influenced by mechanical instabilities of the leaf, when the cellular layers of epidermis and mesophyll grow at different rates. We took a model that had been successfully applied to study phase separation process in alloys, added the interaction with a substrate, and made also the appropriate changes necessary to study the crucial effect of leaf growth. We claim that the properties of biological growth added to the characteristics of the model, explains the formation of a hierarchical structure with well defined statistical properties for different quantities. The results of the statistical analysis are in good agreement with results obtained in actual leaves. Our model explains the existence of abundant closed loops in venation patterns in a natural way. Moreover, some statistical features can be understood analyzing a very simple model of hierarchical division (see Text S1). Our analysis has concentrated in the high order structure of the venation pattern, where it appears isotropic and statistically independent on the particular species that is being studied, and where closed loops are dominant. A further step of investigation would require considering also the first stages of venation growth, where characteristic features of different species appear, and where the existence of closed loops is less universal. We think that it is at this stage where the role of auxin will be critical. In a recent work, Scarpella and collaborators [12] suggested that pre-procambial cells cannot be distinguished by their shape from intact cells at a very early stage of growth. This fact seems to be contrary to the mechanism suggested in this paper, but it is worth emphasizing that our numerically generated patterns have to be interpreted as an indication of the places where the stress is high enough to generate collapsed cells that will eventually differentiate into veins.

A complete and realistic modeling also requires taking into account non-uniform and anisotropic growth, and probably genetic factors [31]. While this is a challenging prospect for future investigation, in its present form our model has some salient interesting features: it provides good statistical agreement of predicted patterns with real ones, and gives a natural explanation for some characteristics of venation patterns, namely the presence of ubiquitous closed loops, which can be accounted for by other models only through the use of very specific hypothesis.

However, it must be stressed that the existence of a instability is an assumption of our modeling, as we do not yet have a confirmation of its existence from a biological point of view. An in situ investigation of this collapse transition along the lines of the experiment made in [26] could help to shed light on the vein pattern formation mechanisms.

## Model

Our main assumption is that vein formation is triggered by the elastic collapse of cells of the mesophyll, growing at a larger rate than the (assumed rigid) epidermis to which they are attached. An appropriate approach would be to describe the mesophyll as an elastic layer with a highly non-linear behavior modeling an irreversible local collapse.

The natural way to theoretically describe the behavior of an elastic layer is by constructing a free energy in terms of the elastic displacement field, **u**. Two main contributions to the free energy should be considered: the elastic interaction between the inner cells and the epidermis, and the energy of the deformed cells that can have two possible internal configurations associated to the intact and collapsed states (see the schematic representation of Figure 8). When this problem is studied in two dimensions the fundamental variable **u** is a two-dimensional vector field. To avoid some technical difficulties that otherwise could appear, instead of studying a non-linear elasticity model directly in terms of **u**, we choose an algorithm in which the elasticity of the cells is assumed to be linear, the non-linear behavior is introduced through an additional field $\Phi$, which is coupled with the elasticity field through a term of the form $\Phi\nabla\mathbf{u}$. The coupling generates the non-linear behavior of the mesophyll in an effective way. This kind of models was successfully used to study phase separation processes in alloys [32–35]. They are described by continuum (differential) equations, and thus the cellular structure of the biological tissues is not considered in detail.

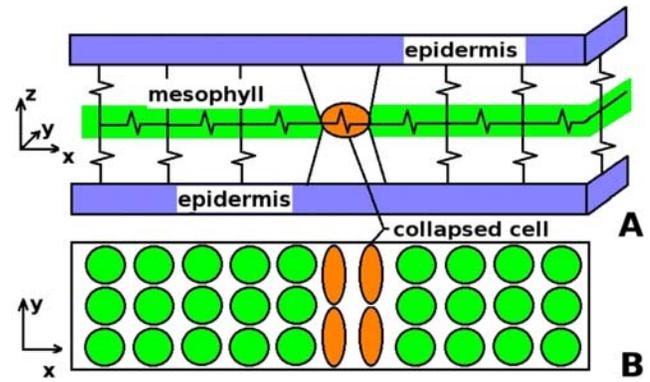

**Figure 8. Schematic representations.** (A) Mechanical analogy. Elastic stresses are accounted for by the springs indicated. Horizontal springs represent the cells of the mesophyll, and its deviation from its equilibrium length is a measure of the deformation energy of the cell. Vertical interlayer springs account for the interaction between mesophyll and epidermis. We suppose that the epidermis grows at a lower rate than the mesophyll, and thus the mismatch between layers will increase with time. A collapsed cell in this schema is represented by a horizontal spring suffering a stress higher than its elastic limit. Once this threshold is reached, the spring has a permanent deformation. (B) Representation of the mesophyll layer with a group of cells in the collapsed state. Note that the initial three-dimensional problem was reduced to two dimensions, as we only describe the intermediate plane where horizontal springs lie.
doi:10.1371/journal.pcbi.1000055.g008

A free energy in terms of the elastic displacement field **u** in the plane of the leaf and the additional phase field $\Phi$, is introduced in the form:

$$F(\Phi, \mathbf{u}) = \int d\mathbf{r}\left[f_0(\Phi) + (C/2)|\nabla\Phi|^2 + \alpha\Phi\nabla.\mathbf{u} + f_{el}(\mathbf{u}) + (\gamma/2)\mathbf{u}^2\right]. \quad (1)$$

Here, $f_0$ is a Ginzburg-Landau local free energy for $\Phi$ that has two different minima, representing the intact and collapsed states:

$$f_0(\Phi) = -(1/2)r_0\Phi^2 + (1/4)s_0\Phi^4.$$

A regularization term proportional to $|\nabla\Phi|^2$ is included to obtain smooth profiles of the fields by penalizing rapid spatial variations of $\Phi$. It is introduced to make the behavior of the system almost isotropic and independent of the underlying numerical lattice. This term is also useful because allows the simulation of a continuous growth through the rescaling of the parameters, as will be explained later.

The parameter $\alpha$ is a measure of the coupling between the fields $\Phi$ and **u**.

The term $f_{el}$ is the usual elastic free energy density in the reference state in which $\Phi = 0$, expressed in terms of the bulk and shear moduli, $K$ and $\mu$, and the displacement field **u**:

$$f_{el}(\mathbf{u}) = (K/2)(\nabla.\mathbf{u})^2 + (\mu/4)\Sigma_{i,j}\left[(\partial u_j/\partial x_i) + (\partial u_i/\partial x_j) - (2/3)\delta_{ij}\nabla.\mathbf{u}\right]^2.$$

We consider the bulk modulus $K$ as constant. However, in order to obtain collapsed regions that can be tentatively associated to growing veins it has to be assumed that the elastic properties of collapsed cells correspond to a lower volume and lower shear modulus than the intact cells (see the morphologies observed in [29]). Thus, the shear modulus $\mu$ will depend on whether the medium is in the collapsed or intact state:

$$\mu = \mu_0 + \mu_1 \Phi \qquad (2)$$

As we said, due to the $f_0$ term, the field $\Phi$ has two preferred values $\Phi_{\pm} = \pm(r_0/s_0)^{1/2}$. When these values are introduced in Equations 1 and 2 they define two different elastic states with different density and shear modulus, representing the intact and collapsed states of the cells in our model. The fact that the variable $\Phi$ is continuous, however, guarantees the possibility of a smooth transition between these states.

The only difference between these expressions and those in the works [32,33,35] is the presence, in our model, of a term proportional to $\gamma$ giving a perfectly harmonic, elastic interaction to a rigid layer that represents the epidermis. Although there are actually two epidermis layers, we suppose their roles are equivalent and thus a single substrate layer is considered in the model. As the growing rate of mesophyll is assumed to be larger than the growing rate of the epidermis, compressive stresses into the mesophyll appear to produce the collapse of some parts of it. This situation corresponds formally to an elastic layer expanding with respect to a rigid substrate, a situation that has been recently studied by one of us [29].

A formal transformation in the model should be made before implementation in the computer. If in the free energy of Equation 1 we were able to integrate out the field $\Phi$, we should end up with a non-linear elastic model written completely in terms of the displacement field **u**. However, the approach we follow is the inverse. Through a well documented procedure [34,35], the elastic field **u** is integrated out of the model to first order in $\mu_1$, and an effective model in terms of $\Phi$ is obtained. The new model is non-linear and non-local in $\Phi$, describing in an effective way the non-linear elastic behavior of the system. The free energy takes the form:

$$F(\Phi) = \int d\mathbf{r} \left[ f(\Phi) + (C/2)|\nabla\Phi|^2 + g_E \Phi \left\{ X_{ij}[1/(\nabla^2 - g_L)] \right\} \right]^2 + \alpha A_{kk},$$

where $X_{ij} = \partial_i \partial_j - (\delta_{ij}/2) \nabla^2$, $g_E = \mu_1 \alpha^2/L_0^2$, $g_L = \gamma/L_0$, $L_0 = K+\mu_0$ in 2D, $A_{ij} = \langle \nabla_j u_i \rangle$, and

$$f(\Phi) = f_0(\Phi) - \Phi(\alpha^2/2) \left[ \nabla^2/(L_0\nabla^2 - \gamma) \right] \Phi.$$

At this point, all the information is encoded in the field $\Phi$. In particular, different values of $\Phi$ in different spatial positions will tell whether that portion of the system is in the intact state, or in the collapsed state. The temporal evolution is governed by an equation compatible with a non-conserved order parameter, i.e., $d\Phi/dt = -\delta F/\delta\Phi$. In this way the system tries to adapt dynamically to the external conditions in order to minimize the value of $F$.

The main external condition that drives the evolution of the system is the fact that the leaf is growing. The natural way to model the growth (which mimics most closely the real situation) is to assume that, although the parameters of the model do not change upon growing, the linear dimension of the system $L(t)$ increases in time. We suppose the growth is sufficiently slow that at each moment the system is in mechanical equilibrium. The initial condition for the minimization at time $t+\Delta t$ should be the result of the minimization at time $t$, but stretched by a factor $L(t+\Delta t)/L(t)$. This approach is quite difficult to implement in the simulation, because of the problems that appear in changing the size of the system under temporal evolution. Technically more simple, but fully equivalent to the previous procedure, is to keep the size over which we integrate the equations of the model, but change its parameters in such a way that the same numerical mesh simulates progressively a larger system. This is like saying that we 'zoom out' with the system growth. The scaling parameter that will do such rescaling is called $\eta$, and the growing process is implemented in terms of changes in the parameters as follows. If in Equation 1 we formally change from **r** to $\eta\mathbf{r}$, the only parameters that are rescaled (in addition of an unimportant global rescaling of the free energy) are $C$ and $\gamma$, which become $C/\eta^2$ and $\gamma\eta^2$. This means that changing $C$ and $\gamma$ in this fashion is precisely the way in which the growing process can be simulated. We start the runs with a value of $\eta = 1$, and increase it progressively during the simulation.

Note the scaling effect in the simulations: Decreasing $C$ will produce a sharper interface between intact and collapsed region, which is a reasonable effect as we zoom out with the system growth. In addition, the increase of the substrate interaction will produce the effective increase of compressive stresses in the active layer, and this will trigger the appearance of new collapsed sectors in order to relieve the accumulated elastic energy.

Our modeling is compatible with the hypothesis that when a new vein has been nucleated in an actual leaf, it will continue to grow at the same pace than the rest of the leaf. In particular its thickness should increase with time. In our modeling, due to our zooming out procedure this means that veins must preserve its width during the evolution and newer veins are progressively thinner than older ones. In order to achieve this, we have to avoid that the older (thicker) veins become thinner as the spatial scale in the system is changed. As we said, this implies a kind of irreversibility condition that guarantees that when a new vein was created, it is committed to grow at a fixed rate. The implementation of the irreversibility condition in the model is as follows. We include the condition that $\Phi$ (x,y) in the time step $t+dt$ can not be in the relaxed phase if its value in the previous time step corresponds to the collapsed phase. This is done by defining a threshold value $\Phi_0$, namely, if at a certain stage of the simulation some point has a value $\Phi$ (x,y)$>\Phi_0$, then this point is forced to remain with a value of $\Phi$ at least as large as $\Phi_0$. Our numerical results indicate that the final patterns are reasonably independent on the value of the threshold we use to define each phase. Irreversibility is what stabilizes the existence of thick veins, as can be observed in Figure 9, where we show a typical profile of $\Phi$ for a fixed value of $x$ at two different stages of the growth. In this plot, values of $\Phi$ close to 2 represent the section of a vein, whereas negative values of $\Phi$ are intact sectors. These results where obtained by using a value $\Phi_0 = 2$. Note in the bottom panel how the interface sharpness is greater (because of the increase in the effective $C$) and how the new nucleated veins are significantly thinner.

It is worth emphasizing the effect that the term that was used to generate irreversibility has on the simulations. In the absence of this term, the same parameters which lead to the snapshots of Figures 2 and 3, produce now patterns like that in Figure 10. A lateral wandering and thinning of veins during evolution is clearly observed. As a consequence, the hierarchical structure is completely lost. Note that in actual leaves a mechanism generating a similar kind of irreversibility can be claimed to be present. In fact, once the germ of a vein has been nucleated, all daughter cells are committed to become part of the vein. This is why older veins are thicker and it is an additional ingredient on top of mechanical energy minimization.

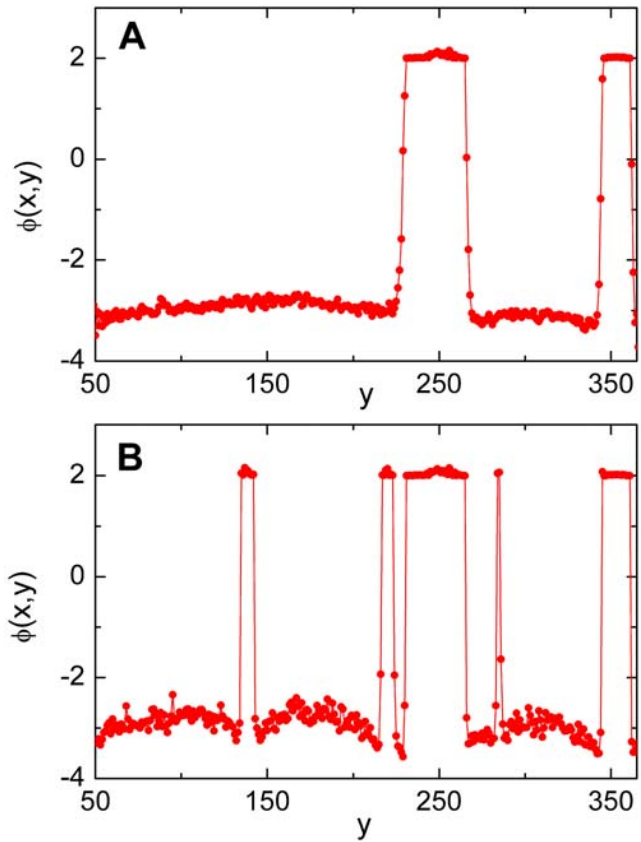

**Figure 9. Profile of the field Φ.** Values of Φ vs. y for a fixed value of x, in two stages of growth. The values of the growing parameter are η = 1.2 (A) and η = 2.4 (B). Positive values of Φ correspond to veins, whereas negatives values are associated to intact tissue.
doi:10.1371/journal.pcbi.1000055.g009

We also include in our model a stochastic noise of small amplitude that helps to nucleate new veins. The evolution equation becomes $d\Phi/dt = -\delta F/\delta\Phi + f^T$, where $f^T$ is a stochastic force with the properties $\langle f_i^T \rangle = 0$ and $\langle f_i^T(t) f_j^T(t') \rangle = 2 k_B T \delta(t-t') \delta_{ij}$. The existence of random noisy effects on the growing of an actual leaf cannot be denied, and then our inclusion of a stochastic term in the evolution equation could be ultimately justified. However, we emphasize that we do not intend to model any precise physical process with this. We only want to include in a simple form the fact that there is some randomness in the nucleation events, which eventually make individual leaves of the same species to differ from one another. In order to be sure that the stochastic term does not introduce systematic spurious effects, we have explored the effect of the noise by applying it in three different ways: 1) a 'static version' in which the noisy term is included only in the initial condition, 2) a dynamic noise as described in the previous paragraph, and 3) an intermediate version, in which a fixed noisy landscape affect the leaf during its evolution. We found that the main characteristics of our patterns as well as its statistical properties are the same in the three cases. Then we present results only for the noisy dynamics, which in addition we consider to be the most realistic one, as fluctuations at the cellular level produced by discrete cellular division events can be considered as some sort of noise during the growing process.

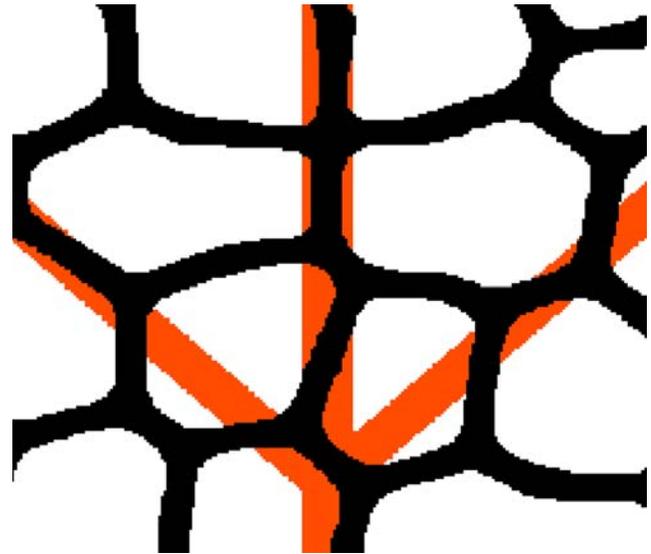

**Figure 10. Patterns without irreversibility.** Result of a simulation with the same parameters as in Figures 2A and 3A, but without the irreversibility condition. Note the lateral wandering and thinning of the veins with respect to the initial condition (shown in a different color). The numerical lattice has 512×512 nodes.
doi:10.1371/journal.pcbi.1000055.g010

## Supporting Information

**Text S1** A minimal model with scale invariance properties. We present here a toy model that has the minimal hierarchical properties we expect to obtain in the full simulation. It may be useful to better appreciate the results of the full modeling.
Found at: doi:10.1371/journal.pcbi.1000055.s001 (0.16 MB PDF)

## Acknowledgments

Early discussions with M. Magnasco are greatly acknowledged.

## Author Contributions

Conceived and designed the experiments: EJ. Performed the experiments: ML. Analyzed the data: ML SB EJ. Contributed reagents/materials/analysis tools: SB. Wrote the paper: ML SB EJ.